# "Never at rest": developing a conceptual framework for definitions of 'force in physics textbooks.

Lars Rikard Stavrum, Berit Bungum, Jonas R. Persson

Programme for Teacher Education, Norwegian University of Science and Technology, NO-7491 Trondheim, Norway

Email: Jonas.persson@ntnu.no

(Dated: June 17, 2015)

## Abstract

*The concept of 'force' is abstract and challenging for many physics students, and many studies have revealed misconceptions that hinder students' understanding and learning in classical physics. One reason for this may be that physics textbooks do not define the concept of force in concise and consistent ways. In the present study, we have investigated how 'force' is defined in physics textbooks used in upper secondary schools in Norway, and present a framework of eight categories for how the concept is defined. The framework is developed inductively from textbooks, and motivated and discussed in light of how the concept has developed historically in physics. Examples from one of the textbooks that constituted the empirical basis for the framework are given. These reveal that textbooks may present students for a multiplicity of definitions of 'force' and hence contribute to students' challenges in learning and using the concept.*

# Introduction

It is well known that understanding the basic concepts and principles of Newtonian mechanics is challenging for students, even after several years of learning physics. Numerous studies have documented students' misconceptions in basic mechanical concepts and principles (see e.g. Alonzo & Steedle, 2009; Coelho, 2012; Viennot, 1979; Warren, 1979). Even if mechanics may appear as a very concrete strand of physics, students here encounter one of the most abstract and in a way difficult concept in physics, namely the concept of force. Consequences of an acting force may be very visible, and students have many experiences of forces in their everyday life, by feeling forces working on their body, or observing the results of a force working on an object. Still, we have no way of observing or experiencing 'force' as such (Carson & Rowlands, 2005). As Gamble (1989) has stated: *"Force is an idea: it is not a concrete object such as an egg or a cell."* (p. 79). This abstract nature of the concept clearly contributes to conceptual challenges in teaching and learning about forces.

Many have suggested that one reason for students' problems is that the term "force" has different meanings and connotations in everyday language than in physics (see e.g. Halloun & Hestenes, 1985). What is less attended to is that the term also holds different meanings in the ways physicists talk about forces, in how it has developed historically and in how it is represented in textbooks and taught to students. This variation may be just as accountable for students' challenges in grasping the meaning of force as is their everyday language use. So far, research in physics education has too little degree investigated how inconsistencies in language use within physics teaching may be a cause for students' problems in developing a robust, scientific conception of force.

In this article, we present a conceptual framework for the variation in how the concept of 'force' is described in physics textbooks and how the different conceptualizations are motivated from the history of physics. Textbook authors are usually well aware of common misconceptions among students, and attempt to counteract them, sometimes by addressing them directly. However, 'force' is used in physics in a number of contexts and with a variety of purposes, and not necessarily in a consistent way across the various contexts it appears in textbooks.

Coelho (2010) has identified three main types of definitions of force:

(i) Force is the cause of acceleration

(ii) Force is defined by the fundamental equations of dynamics, and

(iii) Force is obtained by the connection between force and the effort felt by pulling or pushing an object.

These are used as a starting point for the framework, and categories are extended and refined based on an inductive analysis of four physics textbooks used in lower secondary school in Norway. Even if the categories are based on the textbooks, we present them in a generalized way motivated by fundamental principles in physics and their historical sources. Thereafter, we give examples of how the categories are represented in one of the textbooks that the analysis is based on. This way, we attempt to show how the framework is functional beyond the context of the specific Norwegian textbooks that form the empirical basis for the study.

## Students' challenges in understanding force

For several decades, constructivism was a common conceptual frame for science teaching and research (see Driver, Asoko, Leach, Scott, & Mortimer, 1994). This led to a numerous studies on students' conceptions about natural phenomena (see the bibliography by Duit, 2007), and how to use these in teaching in order to foster conceptual change among students (e.g. Bryce & MacMillan, 2005; Dekkers & Thijs, 1998; Strike & Posner, 1982). A considerable number of studies have been conducted within basic mechanics, and according to Carson and Rowlands (2005), conceptions of force constitute the dominant theme in the misconceptions' literature. Among the widely used instruments for probing students' conceptions about force is The Force Concept Inventory (FCI, see Hestenes, Wells, & Swackhamer, 1992). This instrument, with carefully constructed multiple-choice questions, has been used in schools and universities worldwide for many years, often to access the achievements of teaching methods in mechanics. Results show that the Newtonian understanding of force remains a challenge for students across national contexts and levels of study (see e.g. Caballero et al., 2012; Fulmer, Liang, & Liu, 2014; Savinainen & Scott, 2002). Students' conceptions are often found to mirror outdated ideas in physics. Many students adhere to Aristotelian thinking rather than Newtonian mechanics, despite having studied physics for several years with successful results in terms of marks and credits. They tend to associate force with movement, understand force as something an object carries ("impetus") and will be used up during movement, and anticipate that large objects are acting with larger force than small objects in an interaction (see also Alonzo & Steedle, 2009 for an overview of student conceptions).

Many studies of misconceptions in mechanics and other areas suggest that they form part of students' alternative frameworks, which needs to be challenged and replaced by more scientific ways of thinking. However, as Graham, Berry & Rowlands (2012) have argued, misconceptions are not necessarily held by students prior to teaching, but might just as well be a result of the teaching process. Therefore, studies of how textbooks present the concept of force are important in order to understand how they may contribute to students' inconsistent or flawed conceptions.

## Developing the framework

The empirical material in this study consists of four textbooks for physics in upper secondary school in Norway, and the research is undertaken as an inductive content analysis. Physics textbooks are highly multimodal, and make use of text as well as images, equations, calculations and graphs. The books also contain examples and exercises in addition to the text presenting and explaining the physics content to the learner. All these modes and components of the textbooks are taken into account in the analysis, but with focus on the main running text.

Every paragraph that concerns force in the four textbooks is carefully analysed with regards to how it defines and describes what a force is. As a starting point, we have used the categories developed by Coelho (2010) and described in the foregoing. The analysis demonstrated that these definitions are too coarse and fail to take the gradual development of the concept in textbooks into account. New categories have subsequently been added, refined and adjusted in order to make the framework map the different descriptions of force found in the textbooks as precise as possible and with as few categories as possible. Some categories are divided in subcategories in order to reflect that the same main idea is expressed in different ways. This resulted in a framework of eight categories of definitions of force.

## Definitions of force: Categories

Starting with the three categories of Coelho (2010), it became clear from the textbook analysis that the first category "Force is the cause of acceleration" had to be divided into three categories. This is also consistent with the historical development of the concept. The fundamental equations of dynamics can also be divided into two different categories based on formulations of Newton's second law. Coelho's third category was divided into two categories, one based on Newton's third law and the second on a tactile definition using the Agent-Receiver idea (push-pull). An additional category based on the property of a force to perform work has also been defined, as this description is common in textbooks. The framework thus resulted in eight categories that will be described in the following.

## 1. Effect as property

As force gives rise to an effect, the first category focuses on the effect as a property of force. That is, we have a force if we can observe the effect of the force interacting on an object. It is the effect of the force we observe, and we use this to determine that a force is present. A change in an object's shape (deformation) or a change in motion is a sign that a force is present. In this category it is the effect that works as the definition of force. This can, however, be formulated in slightly different ways, for example *"A force can change the speed of an object, and a force can change form of an object"* and *"The effect of an acting force on an object is that the object changes speed or is deformed".* These two statements seem to have just minor semantic differences, but it is an essential difference in that the latter puts emphasis on the effect while the former focus on the cause.

## 2. Force and motion

As many see force as being the cause of motion, the second category focus on the relationship between force and motion. Note that the definition is based on motion and not acceleration, which has been put in a different category. As there are different ways to introduce force in relation to motion this category may be divided into several sub-categories, and we propose four sub-categories. The order of the sub-categories is of importance as they represent a progressing refinement of the concept.

## 2a) Force as an explanation of motion

The origin of the sub-category can be found in Aristotelean physics, where a force is required to maintain a motion. But one must note that it is not necessarily this idea that lies behind the presentations. Considering that motion, regardless of how it is defined, can be observed and explained by forces, which has in some way caused the motion. This sub-category is, however, based on the absence of causality between force and motion.

## 2b) Force as the cause of motion

As the previous sub-category lacked causality, we defined a sub-category where the causality, a force acting causes a motion, is important. This makes a distinction towards the previous category.

## 2c) Force determines motion

The next sub-category takes a further step towards a more complete description. Here the force is defined as being the direct cause of motion and determines the motion.

## 2d) Force can change motion

In this sub-category the concept of force in developed further as being able to change motion. From the first sub-category where the connection between force and motion was established, the second where force as the cause and third determining motion, the refinement now gives the change of motion as the main point.

## 3. Force is the cause of acceleration

This category is based on definitions of force as the cause of acceleration, and gives a clear connection between force and acceleration. It is noteworthy that it can be represented in different forms, as text or a written mathematical formula. (Coelho, 2010, p. 103) writes that this definition is the most common in textbooks:

> *"In a sample of about a hundred textbooks, it was found that "force is the cause of acceleration" is the most common definition of force."*

## 4. Force and Newton's second law

Having specified a category where a force being the cause of acceleration, there is also need for a category based on Newton's second law. We specify this category in the sense that a mathematical formulation is necessary in order to distinguish it from the "Force is the cause of acceleration" category.

Included are the different representations of Newton's second law. As a sum over forces in vector form or scalar form:

$$\sum \vec{F} = m\vec{a}$$

$$\sum F = ma$$

In most cases, the first presented definition is in the scalar form and for one force only:

$$F = ma$$

It is also possible to present Newton's second law from an acceleration point of view, especially when treating motion without introducing forces first (kinematics) and then introduce forces (dynamics):

$$a = \frac{F}{m}$$

The expression where mass (inertia) is the central tenet can also be included in this category through:

$$m = \frac{F}{a}$$

The basis for this category is Newton's second law with a formulation including acceleration, force and mass.

## 5. Force and momentum

In the previous category Newton's second law served as the basis. It should be noted that Newton did not use this formulation in Principia, but based his formulation on momentum. This category uses the original formulation with the definition of force based on the change in momentum:

$$\sum \vec{F} = \frac{d\vec{p}}{dt}$$

which is the more general formulation.

## 6. Force and Newton's third law

This category includes Newton's third law (N3) as the basis for a definition of force. There exist different ways to formulate N3, but all of these can be used for the definition of force in this category.

Textbooks often refer to force as "interaction". However, the explicit referring might not be to the exact term "interaction", why one have to be aware of the use of different terms, in as much as one does not necessarily describe force as an "interaction".

It should be noted that N3, unlike N1 and N2, in principle does not involve a quantity of motion, yielding that definitions of force in this category often do not have a direct relation between force and motion. N3 may in many ways be said to be "book of rules" for how forces work, and these rules define a force.

Brown (1989) presents five ideas, point by point, as particularly important in conjunction with Newton's 3rd law (N3):

 (i) A body cannot *experience* a force in isolation. There cannot be a force on a body A without a second body B to exert the force.

(ii) Closely related to the above point is the fact that A cannot *exert* a force in isolation. A cannot exert a force unless there is another body B to exert a force on A. We then say that A and B are *interacting*. (Thus, for example, it is incorrect to say that an astronaut punching empty space with his fist is exerting a force since there is nothing exerting a force back against his fist.) The attractive or repulsive force between two bodies arises as a result of the action of the two bodies *on each other* because they are either in contact or experience a force between them acting at a distance.

(iii) At all moments of time the force A exerts on B is of exactly the same magnitude as the force B exerts on A.

(iv) An important implication of the above point is that neither force precedes the other force. Even though one body might be more 'active' than the other body and thus might seem to initiate the interaction (e.g. a bowling ball striking a pin), the force body A exerts on body B is always simultaneous with the force B exerts on A.

(v) In the interaction of A with B, the force A exerts on B is in a direction exactly opposite to the direction of the force which B exerts on A.

These points serve as tools for identifying the category in textbooks.

## 7. Force and work

This category defines a force in relation to work. More specifically, force is defined through the property as being able to do work, hence based on how doing work is defined. In engineering, it is common to define the work W, in the following way:

If a force, $\vec{F}$ is acting on a body such that the body moves a distance, $\vec{d}$, from the starting point, we say that the force has performed a work, $W = \vec{F} \cdot \vec{d}$.

This way, energy is coupled to the concept of force in this category.

## 8. Push-Pull

In the category "Push-Pull", force is 'something' that either 'push' or 'pull' a body. Any force is thus described as a "push" or a "pull".

The "Push-Pull" approach to a force is therefore a definition that builds on the characteristics (properties) of a force. A force can "push" on or "pull" a body, and that this property is observable in as a body can accelerate (change motion) via a "pull" or a "push". "Push-pull" indicates that force *is* a "drag" or a "push", which does not necessarily mean - but does not rule out - that the force must be associated with an action that involves a "push" or a "pull." In this category one can discuss force acting on a body, in the same way as in other categories, but the force is identified as a "pull" or a "push". Even if a force can be described as a "push" or a "pull" this does not necessarily imply what a force 'is'.

# Summary

The categories for definitions of force can be summarized as in table 1.

*Table 1. Categories for definition of force.*

| Category | Description |
|---|---|
| 1. Effect as property | The *effect* of a force serves as the definition of what a force is. |
| 2. Force and motion<br><br>2a) Force as an explanation of motion<br><br>2b) Force as the cause of motion<br><br>2c) Force determines motion<br><br>2d) Force can change motion | The concept of a force is explained through its relation to motion. Multiple relations to motion exists. |
| 3. Force is the cause of acceleration | Acceleration is a consequence of a force acting. |
| 4. Force and Newton's second law | Force is defined through Newton's second law. |
| 5. Force and momentum | Force is defined using the quantity of momentum, and Newton's second law. |
| 6. Force and Newton's third law | The concept of force as described in Newton's third law. |
| 7. Force and work | Force defined using the "ability" of a force to do physical work. |
| 8. Push-Pull | A force is either a "push" or a "pull" (on a body). All forces can be regarded as "push" or "pull". |

## Discussion and motivation of the categories

In the previous section we presented the categories in the framework out of a physical point of view. In this section will we focus on why we defined the categories in this way, with a background and motivation. An apparent overlap between categories will we also be addressed.

Many of the categories have a distinct physical background with different definitions of a force supporting different teaching approaches. Therefore, we have tried to keep the categories as context-free as possible, as to be applicable in other areas of physics besides dynamics. The primary aim has been to build a framework, and we have not evaluated any categories as how and when these should be used in teaching. Even if there exist some progression in sophistication of the categories, the order in which they are presented in textbooks may differ or some may be excluded.

### Newton's first law as a category?

In the previous section Newton's second and third laws have been used in two categories, while the first law is not utilised directly in any category. One reason for this is that the first law describes what happens in absents of forces, that it does not speak of forces as such. It may be argued that it is possible to say a lot about forces while they are absent and thus cause no effect, and compare this with situations where forces are present. The first law states in a way that if you get an effect, change in motion, there must be a force. However, this interpretation is not complete as we shall see.

Newton presented the first law in 1687 as (see Newton, Cohen, & Whitman, 1687, 1999):

*"Every body perseveres in its state of being at rest or of moving uniformly straight forward, except insofar as it is compelled to change its state by forces impressed."*

(p. 416)

One might consider the first law as a combination of two definitions «Definition 3» and «Definition 4»:

*"Inherent force of matter is the power of resisting by which every body, so far as it is able, persevers in its state either of resting or of moving uniformly straight forward."*

(Newton, Cohen, & Whitman, 1999, p. 404)

*"Impressed force is the action exerted on a body to change its state either of resting or of moving uniformly straight forward."*

(Newton, et al., 1687, 1999, p. 405).

With the definitions Newton talks about two "forces" though we would consider the "inherent force" as "inertia", and the "impressed force" as an external force on a body.

Coelho (2010) addresses in his paper Newton's first law from a historical perspective and how the law is presented in textbooks. He points at a problematic aspect of the law, namely that it is not possible to prove it in an experiment, as situation total free from external forces can not be constructed. This opens up for an interpretation that the first law is a special case of the second law. When no forces are present, the acceleration is zero, and therefore any motion will remain constant. This interpretation can be found in a number of textbooks, as pointed out by Coelho (2010). We can find this for example in the widely used textbook University Physics (Young & Freedman, 2012) where the first law is presented this way:

*"It is important to note that the net force is what matters in Newton's first law. For example, a physics book at rest on a horizontal tabletop has two forces acting on it [ ... ] The upward push of the surface is just as great as the downward pull of gravity, so the net force acting on the book (that is, the vector sum of the two forces) is zero. In agreement with Newton's first law, if the book is at rest on the tabletop, it remains at rest. [ ... ] When a body is either at rest or moving with constant velocity (in a straight line with constant speed), we say that the body is in equilibrium. For a body to be in equilibrium, it must be acted on by no forces, or by several forces such that their vector sum – that is, the net force - is zero [ ... ]*

(Young & Freedman, 2012, p. 112).

Similar presentations can be found in other textbooks. Even if the first law is not explicitly identified as a special case of the second law, it is done indirectly through the reference to a zero net force.

Whether this is problematic or not depends in which context and how one chooses to formulate the first law. As Coelho (2010) points out, the first law defines a reference-state or reference-motion, when no external forces are present. The first law then works as an axiom, necessary in explaining that if there is an effect there must be a cause. In other words, the first law must be there in order to explain in an unambiguous way that if there is an effect due to a force there must be a force in connection with the laws of gravity. In terms of our categories, this is an inverted version of the category, Force can change motion, as it couples a change in motion to the existence of a force.

This line of thought might not appear in textbooks, but gives the background why it is difficult to include a category based in the first law. Since the first law is based on basic assumptions of the existence and absence of forces it does not describe what a force is, just if it is present or not.

## Effect as property

This category is mainly based on experience and empirical observations. From observing how it is possible to change the motion of a body, one might obtain an understanding of why a change happens. This understanding is not necessarily based on knowledge of physics or the physical concept of force, but rather on visual or tactile observations in everyday life.
The category has the obtained effect as the main characteristic, the observed effect is important. If we look at two common ways to talk about forces, namely

 *" A force can change the speed of an object, and a force can change form of an object"*
and

*" The effect of an acting force on an object is that the object changes speed or is deformed."*

we observe that they are not equivalent. The semantic differences lies in the use of the words "can" and "or", which may have different connotations in different languages. Instead of stating that a force *can* change the speed, it is possible to say that a force *may* change the speed, or that a force *will* change the speed. This contextual factor that is important in how this category is analysed. It can be interpreted as we have two dimensions in this category, which in a sense is true but this will depend on the semantic context of the language.

In our case, where the language is Norwegian, we have two dimensions. One definition based on the possibility that a force *may* give rise to the effect and one where a force *will* give rise to an effect, where the former being less precise within this category as we can apply or experience a force without an apparent effect.

## Force and motion

The need for an agent or force (mover) acting in order to have or uphold motion is an ancient idea, and has been part of the framework of many natural philosophers. Aristotle describes a concept of a universal cause, prime mover or primum movens, that is the ultimate cause of motion in the universe (Jammer, 1999). The idea of a prime mover is also present in the work of Newton, but to some extent less dominant. The idea that there must be a cause to motion was considered systematic in the work of Aristotle, but the concept of force as we know it today was non-existent, however the word was used. Aristotle identified "force" as the cause of motion, unlike objects being at rest, but that "force" is not what we use today. We will not go into detail on this beyond stating that the idea that force and motion are connected forms the major idea of this category.

One must note that by motion we do not necessarily mean acceleration. The concept of motion is not well defined and will imply a number of alternatives. As we use the term motion "as is", a clear definition of motion will set boundaries and give rise to additional categories or sub-categories. This category is also based on a textual description, without a mathematical formalism.

As there are different ways to "define" force in connection with motion we identified four sub-categories: Force as an explanation of motion, Force as the cause of motion, Force determines motion and Force can change motion.

The force as such is the same but it is the relation to the description of motion that changes. The difference is not entirely semantic but shows an increasing degree of sophistication in how the motion is affected by a force.

## Force is the cause of acceleration

A further sophistication from the previous category is to couple force and acceleration, and to say that force is the cause of acceleration. One should be aware that it is possible to introduce a semantic distinction between "force is the cause of acceleration" and "force causes acceleration". However, we will not divide this category into sub-categories, but rather use this category as identified by Coelho (2010, 2012). He noted that the definition (category) «force is the cause of acceleration» is the most common in physics textbooks.

Taking into account that acceleration is a more precise defined quantity than motion, in addition to being measurable, one might argue that this category is a sub-category to "Force and motion". It could also be a sub-category to "Force and Newton's second law". However, as noted above, we make use of the clearer definition (acceleration), which gives a clear distinction to the former categories. We also lack a clear mathematical definition as in Newton's second law, which therefore is on a higher level of sophistication.

With clearer definitions, it is more difficult to introduce sub-categories as in the previous category. *Force as an explanation of acceleration* seems to be distinct from *Force is the cause of acceleration*, but since the cause is an explanation they are in all practical cases equal.

## Force and Newton's second law

The basis for this category is Newton's second law. However, it is not the general form of Newton's second law that is used in this category but the different acceleration-based forms. This is partly motivated by the fact that it is this form that is most common in physics textbooks, but also in the difference in sophistication between the acceleration form and general form of Newton's second law.

From Principia the second law is stated as:

> *"A change in motion is proportional to the motive force impressed and takes place along the straight line in which that force is impressed."*

(Newton, et al., 1687, 1999, p. 416),

where «motion», is defined as «velocity and the quantity of matter jointly» (p. 404), that is «momentum». We will use this definition in the next category *Force and momentum*.

This gives Newton's second law as a change in momentum per time unit:

$$\sum \vec{F} = \frac{d\vec{p}}{dt}$$

In the case of constant mass, this gives:

$$\sum \vec{F} = \frac{dm\vec{v}}{dt} = m \frac{d\vec{v}}{dt} = m\vec{a}$$

That is the acceleration form of Newton's second law, which is the basis for this category.

One might consider this as a sub-category to *Effect as property*, as we define a force by its effect, in this case acceleration. However, as noted above, the categories follow an increasing sophistication, that is, we include more precise definitions and refined mathematical expressions.

In other words just by formulating Newton's second law in the acceleration form, one is within this category. As indicated by Roche (2006), it is possible to have a definition of force based on the product of mass and acceleration, in other words Newton's second law.

## Force and Momentum

The previous category "Force and Newton's second law" is based on Newton's second law defined with acceleration. In this category, we also use Newton's second law, but now defined using momentum. It might be possible to combine both categories into a Newton's second law category, but since the quantities differ in the definition of Newton's second law, we have chosen to have two categories. In addition, one can find both categories in the way university textbooks present Newton's second law, for example Alonso & Finn Fundamental University Physics (momentum) versus Sears & Zemanskys University Physics (acceleration).Stating Newton's second law as

$$\sum \vec{F} = \frac{d\vec{p}}{dt}$$

relates force to momentum, or rather to the change in momentum (impulse). As with acceleration, momentum is a quantity that is clearly defined and can be measured. In this case mass is included in the definition of momentum.

In Principia, Definition 2, Newton defines the concept of motion, which we recognize as

momentum:

> *"Quantity of motion is a measure of motion that arises from the velocity and the*
>
> *quantity of matter jointly."*

> (Newton, et al., 1687, 1999, p. 404).

This definition is used in the second law:

> *"A change in motion is proportional to the motive force impressed and takes place*
>
> *along the straight line in which that force is impressed"*

> (p. 416).

Cohen stresses in the guide to Principia (Newton, et al., 1687, 1999) that Newton actually

means motion as in definition 2, that is momentum, and not the loose definition of motion

used in the previous definitions:

> *"Def. 2 states that the measure of motion adopted by Newton is the one arising from*
>
> *the mass and velocity, our momentum. Although Newton does not say so specifically [*
>
> *... ] it is this quantity of motion that he means when, as is often the case, he writes*
>
> *simply of 'motion'. For example, in law 2, Newton writes that a 'change in motion' is*
>
> *proportional to the motive force.' Here he means 'change in the quantity of motion'*
>
> *or, in our terminology, change in momentum."*

The form of Newton's second law used in the category is not equivalent with the form in Principia. We use a continuous force while Newton included the possibility for an instantaneous force, an impulse or impulsive force.

If one changes the formulation of Newton's second law to «The rate of change in motion ... » (Newton, et al., 1687, 1999, pp. 111-112), one obtains the form used in this category.

An important aspect of using the momentum-form of Newton's second law is that momentum is a conserved quantity, while acceleration is not conserved.

## Force and Newton's third law

This category uses Newton's third law (N3) as the basis for a definition of force, which is force as an interaction. In this case, force is not defined by the effect of forces, such as acceleration, but rather how forces interact. This may be considered as more abstract than other categories, as force is not defined through a relation. However, by using Newton's third law as a way to describe the action of applied forces, the category can be considered to be precise. It is important to include the five points that Brown (1999) describes as important in the definition of the category, in order to emphasise how a definition of force will fit into this category.

In Principia, the third law is stated as:

*"To any action there is always an opposite and equal reaction; in other words, the actions of two bodies upon each other are always equal and always opposite in direction"*

(Newton, et al., 1687, 1999, p. 417).

Newton's third law is presented in the same form in modern literature; however the emphasis is slightly different. In this category, the emphasis is on the interaction between two objects, that is the force acting between them. This is similar to the definition in the category "Effect as property", where the effect of the force defines force. Newton's third law does not describe a specific effect, but rather how forces interact. This category can hence be seen as an abstract version of the Push-Pull category, where the tactile experience is important.

## Force and work

With the progression in sophistication of how forces are presented, force is coupled to other concepts not directly associated with motion. This category is also related to energy, through the concept of work. We have chosen to avoid a more sophisticated subcategory, and include all references to work and energy to a category of its own.

In dynamics, work is defined in terms of applied force and displacement:

$$W = \vec{F} \cdot d\vec{s}$$

More generally can work be defined through the transfer of energy from one system to another. In this case, we limit the category to the mechanical definition. A force is said to do *work* when it acts on a body so that there is a displacement of the point of application in the direction of the force. Thus, a force does work when it results in movement.

It is the motion through work that defines a force, or rather that motion is defined by the work a force does. As we have used motion in a previous category, it is the use of work in this category that is the hallmark. Since work is closely related to energy, the category also includes the concept of energy.

The kinetic energy of an object of a specific mass, and a speed, can be defined as the work done to accelerate the object from rest to a speed (v). As work is related to the sum of forces acting on an object, force may be included in the definition of kinetic energy. This means that a force may increase (or decrease) the kinetic energy of an object. This implies a possible overlap with other categories, such as "Force and motion" and "Effect as property". It is as specified above the use of the concepts work and energy that defines this category and the higher degree of sophistication compared with the other categories.

## Push-Pull

This category is not based on a concrete context or any form of physical expression or law, and hence hard to define. It is based on descriptions or a model of forces. The use of "push" and "pull" and the concept of force can be found in different textbooks, but also in more general contexts. The "push-pull" category consists of a set of descriptions of force, which are intended to help students understanding the concept of force.

Forces are either a "push" or a "pull". This can make it easier to understand the concept of force, since one is used to push or pull things in everyday life, to get a desired effect. The category might be considered as a tactile category where the experience is important. It is important to note that this category does not limit itself to contact forces, as one can also experience long-range forces such as gravity and magnetic forces.

Even if the category is not primary based on physical laws, it has an important conceptual dimension in physics and the history of physics. The fundamental idea is based on pre-Newtonian mechanics, where philosophers as Aristotle and Rene Descartes, believed that only contact forces were possible. That is, an object could only act upon another object with a force if they are in contact with each other or through a solid mediator (Jammer, 1999). Descartes tried to construct a theory of gravity in order to explain the movement of the planets based on contact forces, since "action at a distance" didn't exist in his philosophy. (Jammer, 1999) describes this in his book:

> *"For, rejecting any possibility of an action at a distance, Descartes constructs his vortex theory to account for the remote heavenly motions. To assume some action at a distance for their explanation would be tantamount, he claims, to endowing material particles with knowledge and making them truly divine, «as if they could be aware, without intermediation, of what happens in places far away from them."*

> (Jammer, 1999, p. 104).

If we look at Newton's concepts of force, we see a difference compared to Descartes. Cohen and Smith (2002) describe the different forces found in Principia:

> *"Centripetal forces differ from percussion and pressure in one notable aspect. Percussion and pressure are the result of some kind of observable physical action. In both, there is a contact of one body with another, typically providing visual evidence* *of a force acting, for example, a billiard ball striking another billiard ball. These are*

*the kinds of force on which the so-called "mechanical philosophy" was built, in*

*particular the philosophy of Descartes, These forces display the principle of matter in*

*contact with other matter to produce or alter a motion"*

(Cohen & Smith, 2002, pp. 62-63).

There are of course more aspects to the force concept in pre-Newtonian mechanics, but in this category the view that only contact forces are possible is essential. In the push-pull category forces may be considered as either a push or a pull between two objects in contact or a tactile experience of force.

Even if the historical background is strong, it has a more everyday approach, as we can see in University Physics (Young & Freedman, 2012):

*"In everyday language, a **force** is a push or a pull. A better definition is that a force is*

*an interaction between two bodies or between a body and its environment."*

(p.108)

This approach refers to force as in the Push-pull category. The category will in that sense be quite superficial, while the other categories will go deeper into physical definitions. This makes this category special, and hard to define.

Coelho (2012) indicates a class of definitions of force that are based on "the effort felt by the pulling and pushing of an object":

*"In contemporary textbooks on mechanics, three kinds of definitions of force were found. [ ... ] In some textbooks, a connection between force and the effort felt by the pulling or pushing of an object has been established."*

(Coelho, 2012, p. 1339).

Several authors indicate how force as a push or a pull, is a common definition. Gamble (1989) writes:

*"The basic idea of force as a pull or a push is developed into the relationship between force and motion, distortion, acceleration, gravity and weight, pressure, work, power, energy and so on."*

(p. 79)

Hart (2002) notes that:

*"It is interesting that most textbook treatments of this topic [Newton's laws of motion] assume that the notion of what a force 'is' is either self-evident or can be easily dealt with by defining force as a push or pull."*

(p. 234)

Hart also argues, based on her experience, that the push pull definition has little value in teaching about forces in classical mechanics:

*"The simplistic definition of force as a 'push or pull' is of little help beyond, perhaps, excluding the metaphorical use of the term»*

(p. 237)."

The category can be summarized as conceptualising force as a "push" or a "pull". Note that it is the force that is defined - not the effect of force.

## Representations of categories: Examples from one physics textbook

In the following, we give examples of how the categories are represented in one of the textbooks that form the empirical basis for the framework. The textbook is adapted to the curriculum for physics (Physics 1) in year 12 in Norwegian upper secondary school.[1]

The physics curriculum consists of five main subject areas named "classical physics", "modern physics", "explaining nature through mathematics", "the young researcher" and "physics and technology". The concept of force appears explicitly mainly within the main subject area *classical physics*, but is also of relevance within other areas.

Competence aims in classical physics for Physics 1 are given in Table 2. As can be seen, the curriculum document gives in itself no definition of what 'force' means, but rather the various context where students should apply it and the calculations and qualitative descriptions they are supposed to master through taking the course.

---

[1] The curriculum is available from www.udir.no

*Table 2. Curriculum aims for classical physics in Physics 1 in the curriculum for Norwegian upper secondary school (from www.udir.no)*

| **Physics 1. Students should be able to:** |
| --- |
| - identify contact forces between objects and gravitational forces on objects, draw force vectors and apply Newton's Three Laws of Motion |
| - give an account of the concepts of energy, work and effect, carry out arithmetic calculations and discuss situations where mechanical energy is conserved |
| - give an account of situations where friction and air resistance mean that the mechanical energy is not conserved, and perform calculations in situations with constant friction |
| - state and discuss the first and second laws of qualitative thermo physics |
| - define the terms current, voltage and resistance, and apply the principles of conservation of charge and energy to simple and branched direct current circuits |
| - define and carry out calculations with the terms frequency, period, wavelength and wave speed, and explain qualitative bending and interference phenomena |

## Textbook examples: The case of "Rom Stoff Tid 1"

In our case we consider the textbook "Rom Stoff Tid Fysikk 1" (RST 1, Jerstad, Sletbak, Grimenes & Renstrøm, 2007). This textbook does not follow the composition of traditional pre-university physics textbooks, as it starts with topics from modern physics, where forces are not mentioned. The chapters 5, 6 and 7, "Motion", "Force and motion" and "Work and energy" respectively, constitute the part of RST 1 covering classical mechanics, and we hence limit our case study of RST 1 to these three chapters.

### Chapter 5: "Motion"

Chapter 5 does not deal directly with forces, but introduces the reader to the central quantities in mechanics: displacement, position, speed and acceleration. These are presented as scalar quantities, and towards the end of the chapter it is mentioned that the quantities are in fact vector quantities. One can thus summarize the chapter as an introduction to basic quantities, enabling the reader to describe motion without the introduction of force.

## Chapter 6: "Force and motion"

This chapter introduces the reader to forces in the domain of classical mechanics.

In the introduction, "force" is established as an important concept in physics. The reader is given a brief explanation as to how the chapters 5 and 6 are connected: "In the previous chapter the subject of study was how to describe and measure motion, without considering the forces behind the motion. In this chapter however, force in relation to motion is the subject of study." (RST 1, p. 105). The implication is that the quantities introduced in chapter 5, position, speed (velocity) and acceleration, are descriptive quantities in regards to motion. Force on the other hand seems to be more of an explanatory quantity through its relation to motion. This is an indication that the "Force and motion"-category is represented, but perhaps not yet a specific sub-category, as the phrase "force in relation to motion" is not a clear statement.

The first statement of the chapter itself is: "A force is an interaction between *two bodies*." (RST 1, p. 106). Thus, the notion of an "interaction" appears already in the very first sentence of the chapter. This first statement might be said to serve as a definition (though, it is not highlighted as a definition, as are other laws and definitions in the textbook) of what a force is – an *interaction*. The use of the term "interaction", and the fact that RST 1 specifies that a force involves two bodies, indicates the presence of "Force and Newton's third law" category.

This initial statement is immediately followed by an explanation: "By this we mean that two bodies can interact with each other in such a way that their velocity and/or shape are changed." (p. 106, RST 1). This can be categorized as being a description of "Effect as property" as the textbook says something about what a force has the ability to do[2]. The focus is on the fact that a force actually has an effect or an "impact". Then follows an example and a list of three points summarizing the content of these initial statements as to the properties of a force: "1. A force always acts *from* one body *on* another. Thus, there are always *two* bodies involved. 2. A body being acted upon by a force from another body, always act with a force on the other body. 3. A force can[3] change the speed (velocity) and/or the shape of a body." (p. 106, RST 1). The second statement can be categorized as "Force and Newton's third law".

 "Interaction between two objects: Newton's third law" is the following sub-chapter. It is interesting to note that the textbook does not to present Newton's laws in "numerical" order. That is, Newton's third law is introduced before the first and second law. RST 1 justifies this by stating that "Newton's third law is important to gain a better understanding of the concept of force, before we introduce the other of Newton's laws" (p. 110, RST 1). This sub-chapter focuses on the fact that all forces come in pairs, as described in Newton's third law, placing the presentation in the "Force and Newton's third law" category.

The introduction of Newton's first and second laws is given in a dedicated sub-chapter called "The relation between forces and motion: Newton's first and second law" (p. 114, RST 1).

Newton's first law is stated as: "An object will remain in a state of rest or in a state of rectilinear motion with constant speed unless forces compel it to change this state" (p. 114, RST 1). While it is obvious that force appears in the category "Force and motion" through the statement of Newton's first law, it is worth noting the choice of words. As it appears, a force has the ability to change the state of motion of an object, which is the sub-category "Force can change motion".

---

[2] This refers to the translation of the Norwegian word "kan", which we chose to translate with the English "can".
[3] In this particular context, we have chosen to translate the Norwegian word "kan" with the English "can". One could perhaps have chosen the English "may" as a suitable translation. This is a matter of interpretation, as is the case of every piece of translated text presented in this article. The translations in this article are the authors own.

Newton's second law is stated, using both the mathematical expression as well as textual explanations. It is stated that the speed of a body will change when the sum of the forces acting is non-zero. That is, force is related to motion through the mathematical expression of Newton's second law and appears in the category "Force and Newton's second law". An example following Newton's second law consider the act of pushing a car so it starts (p. 117, RST 1), where the "pushing", together with all other forces action on the car (the sum of the forces) is identified as the cause of the acceleration of the car, thus linking the category "Force is the cause of acceleration" to the more refined category "Force and Newton's second law".

As we have seen, the concepts of force appear within multiple categories in this chapter of the RST 1 textbook. Table 3 displays the different categories that appear, and in which order. It is important to note that the textbook include five out of eight categories, indicating a large diversity in how the concept of force is used in textbooks. Though the presence of the different categories varies, as indicated in the above analysis, the concept of force is given five substantially different "definitions" or explanations, which the reader has to take into consideration.

*Table 3. Categories of force represented in chapter 6 of RST 1. The "Order" column shows the order in which the different categories appear in this chapter. Only the first appearance constitutes the order of the category.*

| Category | Order of appearance |
|---|---|
| "Effect as property" | 2 |
| "Force and motion" | 1 |
| "Force is cause of accel." | 5 |
| "Force and Newton's 2." | 4 |
| "Force and momentum" | - |
| "Force and Newton's 3." | 3 |
| "Force and work" | - |
| "Push-pull" | - |

## Chapter 7: "Work and energy"

The introduction of this chapter states that if we know all the forces acting upon a body, we can use Newton's second law to calculate how the motion of the body changes (RST 1, p. 131), but that it is also possible using the "energy method" as an alternative, which is a less complex method of calculating the motion of a body.

In introducing the concept of work, the textbook presents a large rock falling from a cliff into the sea. In analysing this situation, it is said that: "[…] gravity pulls it [the block of rock] downwards, ever increasing its speed. We say that the speed of the block of rock is continuously increasing because gravity is doing work on it" (RST 1, p. 132). The sequence presents gravity as a "pulling" force with the ability to do work", that is, both the "pulling" and the "doing work" is associated with a change in the speed of a body. One uses the category "Force is the cause of acceleration" to introduce the concept of "work". In other words this chapter mentions "work" as something a force "does". Following the example with the rock, it is defined that the physical quantity work is the product of force and displacement (RST 1, p. 132). Such a notion of a force can be categorized within the category "Force and work".

The definition of work is refined taking into account that a force might not act in the same direction as the displacement vector.

## Final discussion and conclusion

The framework of definitions of force illustrates that the concept is multifaceted and that definitions can be traced back to different historical sources. Our textbook examples show that students have to deal with this multiplicity even in the same book, often without being made aware of the variability of how the concept of force is defined and described. Rather, textbooks seem to switch between definitions that are appropriate for each purpose depending on the specific topic to be presented. This may in itself be a source of the conceptual problems that many students have with understanding and using the concept appropriate in describing and solving physics problems. It may also cause students to stick to intuitive everyday conceptions, as the various definitions of force they encounter in physics textbooks might be counteractive to each other and students understanding.

Several authors have suggested that including aspects of the history of physics may help students to understand not only how physics has developed as a human product, but also to get a better understanding of the concepts involved (e. g. Coelho, 2010; Gauld, 2014). Historical approaches also have the potential to make students aware of the fact that the principles we refer to as Newton's laws today look quite different from what Isaac Newton originally formulated. This may help them notice and make sense of the various conceptions of force they encounter in textbooks.

More research should be undertaken in order to examine the degree to which the multiplicity of definitions of force is present in physics textbooks in other countries and on other levels of education. The framework presented in this paper represents a constructive tool for analysing and comparing textbooks in this regard. It also has a potential for analysing classroom teaching and student responses about forces. This may contribute to illuminate how different conceptualizations of force, and the variability in itself, may influence students' understanding and their learning processes.